\documentclass[aps,prl,twocolumn,superscriptaddress]{revtex4}

\usepackage{graphicx}
\usepackage{bm}

\begin{document}

\preprint{}

\title{
Unconventional multiband superconductivity with nodes in single-crystalline SrFe$_2$(As$_{0.65}$P$_{0.35}$)$_2$ as seen via $^{31}$P-NMR and specific heat
}

\author{T. Dulguun}
\affiliation{Graduate School of Engineering Science, Osaka University, Osaka 560-8531, Japan}

\author{H. Mukuda}
\email[]{e-mail  address: mukuda@mp.es.osaka-u.ac.jp}
\affiliation{Graduate School of Engineering Science, Osaka University, Osaka 560-8531, Japan}
\affiliation{JST, TRIP (Transformative Research-Project on Iron-Pnictides), Chiyoda, Tokyo 102-0075, Japan}

\author{T. Kobayashi}
\affiliation{Department of Physics, Graduate School of Science, Osaka University, Osaka 560-8531, Japan}

\author{F. Engetsu}
\affiliation{Graduate School of Engineering Science, Osaka University, Osaka 560-8531, Japan}

\author{H. Kinouchi}
\affiliation{Graduate School of Engineering Science, Osaka University, Osaka 560-8531, Japan}

\author{M. Yashima}
\affiliation{Graduate School of Engineering Science, Osaka University, Osaka 560-8531, Japan}
\affiliation{JST, TRIP (Transformative Research-Project on Iron-Pnictides), Chiyoda, Tokyo 102-0075, Japan}

\author{Y. Kitaoka}
\affiliation{Graduate School of Engineering Science, Osaka University, Osaka 560-8531, Japan}

\author{S. Miyasaka}
\author{S. Tajima}
\affiliation{Department of Physics, Graduate School of Science, Osaka University, Osaka 560-8531, Japan}
\affiliation{JST, TRIP (Transformative Research-Project on Iron-Pnictides), Chiyoda, Tokyo 102-0075, Japan}

\date{\today}

\begin{abstract}
We report $^{31}$P-NMR and specific heat measurements on an iron (Fe)-based superconductor SrFe$_2$(As$_{0.65}$P$_{0.35}$)$_2$ with $T_c$=26 K, which have revealed the development of antiferromagnetic correlations in the normal state and the unconventional superconductivity(SC) with nodal gap dominated by the gapless low-lying quasiparticle excitations. 
The results are consistently argued with an unconventional multiband SC state with the gap-size ratio of different bands being significantly large; the large full gaps in $s_{\pm}$-wave state keep $T_c$ high, whereas a small gap with a nodal-structure causes gapless feature under magnetic field. 
The present results will develop an insight into the strong material dependence of SC-gap structure in Fe-based superconductors.

\end{abstract}

\pacs{74.70.Xa, 74.25.Ha, 76.60.-k}

\maketitle


Superconductivity (SC) in the 122 iron-pnictides $M$Fe$_2$As$_2$ ($M$=Ba, Sr or Ca) can be induced by either electron- or hole-doping, an application of pressure or chemical substitution for $M$, Fe, and As. 
The substitution of K$^{+1}$ for Ba$^{+2}$ in (Ba$_{1-x}$K$_{x}$)Fe$_2$As$_2$ (denoted as BaK122) suppresses antiferromagnetism (AFM) and structural transition  through the hole-doping, and induces SC with the relatively high superconducting transition $T_c$=38 K at $x$=0.4~\cite{Rotter}. Extensive experimental works in this compound have revealed the fully-gapped $s_{\pm}$-wave state~\cite{Hashimoto1,Ding,Yashima,Xu,Zheng}. 
The isovalent substitution of P for As in BaFe$_2$(As$_{1-x}$P$_{x}$)$_2$ (denoted as Ba122P) also exhibits the maximum of $T_c\sim$30 K at $x$=0.33, providing a phase diagram similar to BaK122 without changing a carrier density~\cite{Kasahara}. Notably, a nodal-gap structure in the SC state has been reported by various experiments on Ba122P~\cite{Kasahara,NakaiPRB,Hashimoto2,JSKim,Wang,Yamashita}. Relevant with this nodal-gap structure, the $^{31}$P-NMR study on Ba122P~\cite{NakaiPRB} reported the gapless SC with a large fraction of residual quasiparticle density of states (RDOS) $N_{\rm res}$ at the Fermi level ($E_F$) with $N_{\rm res}/N_{0}\sim$0.34, where $N_{0}$ is the DOS at $E_F$ in normal state. 
On the one hand, we note that a large fraction of $N_{\rm res}/N_{0}\sim$0.5 was reported even for the SC state of undoped SrFe$_2$As$_2$(Sr122) under highly hydrostatic pressure\cite{Kitagawa}. It is therefore likely that the presence of RDOS is inherent in the 122 iron-pnictides in which a lattice contraction takes place by either the application of pressure or the chemical substitution of P for As.
At the present, they are not clear which Fermi surfaces (FSs) are responsible for the existence of nodal-gap and why it exists there\cite{Suzuki_Kuroki,Yoshida,Shimojima,Feng,Yamashita}. 

The replacement of either Sr or Ca for Ba site in $M$122P compounds brings about further lattice contraction. 
A phase diagram for SrFe$_2$(As$_{1-x}$P$_{x}$)$_2$ (Sr122P) resembles that for Ba122P\cite{Shi}, but a maximum $T_c$ decreases to 26 K at $x$=0.35, and a maximum $T_c$ for CaFe$_2$(As$_{1-x}$P$_{x}$)$_2$ (Ca122P) does to $T_c$=15 K at $x\sim0.04$ in a narrow SC region\cite{Kasahara2}. 
These results imply that the effect of lattice contraction may cause a reconstruction of FS topologies\cite{Yoshida}, causing the onset of SC to be unfavorable, but it provides us with a good opportunity to investigate an intimate change in electronic state and SC gap structure systematically in the $M$122 iron-pnictides.

In this paper, we present $^{31}$P-NMR results on single crystalline SrFe$_2$(As$_{0.65}$P$_{0.35}$)$_2$ (Sr122P) with $T_c$=26 K, which unravels the development of AFM correlations in the normal state and the unconventional gapless SC with nodes. Specific heat measurement in SC state reveals that the fraction of RDOS at $E_F$ are mostly induced by the external field. 
Although the low-lying quasiparticle excitations are gapless in a small gap even when $H$=0, the fully-gapped $s_{\pm}$-wave state in other bands is not affected significantly. 
Such multiband effect plays a significant role in keeping $T_c$ relatively high in this compound.

\begin{figure}[h]
\centering
\includegraphics[width=7cm]{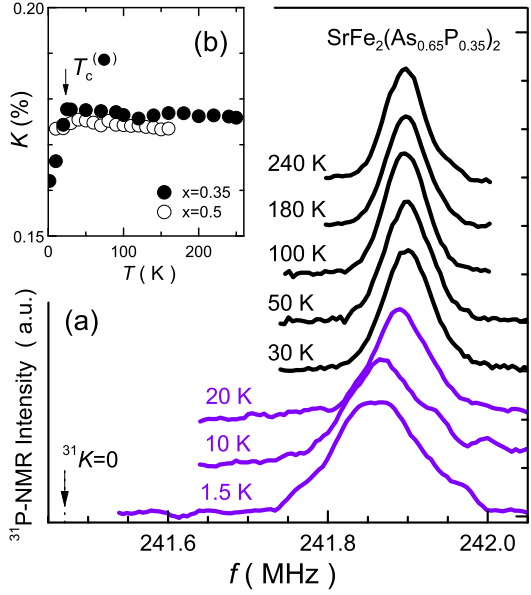}
\caption[]{(Color online) (a) $T$ dependence of $^{31}$P-NMR spectra for Sr122P($x$=0.35) at $H$=14.0088 T. (b) $T$ dependence of Knight shift $^{31}K$, which is $T$-invariant, but decreases below $T_c(H)$=23 K.  The $^{31}K$ for $x$=0.5 is slightly smaller than that of $x$=0.35 in the normal state, which is attributed to the subtle reduction of DOS at $E_F$\cite{NakaiPRL}.} 
\label{NMRspectra}
\end{figure}

Single crystalline samples of SrFe$_2$(As$_{1-x}$P$_{x}$)$_2$ ($x$=0.35 and $x$=0.5) were synthesized by self-flux method\cite{Kasahara,Kobayashi}. 
We used  small pieces of single crystals that are superconducting at $T_{\rm c}$=26 K for $x$=0.35 with a sharp transition width $\Delta T_{\rm c}<$1 K. 
$^{31}$P-NMR measurement for $x$=0.35 was performed at fields $H$=1$\sim$ 14.0 T perpendicular to the $c$-axis using the aligned single crystalline samples. 
The Knight shift $^{31}K$ was measured with respect to a resonance field in H$_3$PO$_4$ ( $^{31}K$=0 in Fig.~\ref{NMRspectra}). 
For comparison, these measurements were also performed on $x$=0.5 with $T_c$=8 K in the overdoped regime.



Figure \ref{NMRspectra} shows a temperature ($T$) dependence of $^{31}$P-NMR spectrum at $H$=14.0088 T for $x$=0.35.
The NMR spectral shape does not change down to $T_c(H)$=23 K with a full-width at half maximum (FWHM) of 55 kHz at $\sim$14 T and 250 K, which corresponds to $\Delta K < 0.02$\%, indicating that the present single crystals are totally homogeneous. Upon cooling below $T_c$, its peak shifts to a low frequency side and the FWHM increases owing to the appearance of SC diamagnetism\cite{NakaiPRB}.
As indicated in the inset of the figure,  $^{31}K$ is nearly $T$-invariant in the normal state, but decreases below $T_c(H)$.
In the $x$=0.5, the FWHM becomes 95 kHz at 100 K, being broader than that of $x$=0.35, due to the higher doping level. The $^{31}K$ in the normal state is smaller than that of $x$=0.35, which is attributed to the subtle reduction of DOS at $E_F$, being in good agreement with the result in Ba122P\cite{NakaiPRL}. 

\begin{figure}[h]
\centering
\includegraphics[width=7.5cm]{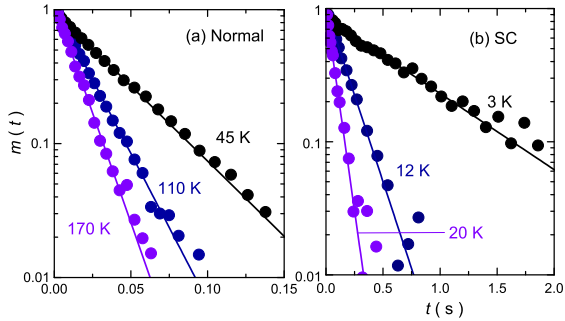}
\caption[]{(Color online) Nuclear spin-lattice relaxation rate $1/T_1$ was obtained from the recovery of nuclear magnetization by fitting with a simple exponential recovery curve of $m(t) = [M_0-M(t)]/M_0 = \exp(-t/T_1)$  in (a) normal and (b) SC state for $x$=0.35 at $H\sim$14 T. }
\label{recovery}
\end{figure}

\begin{figure}[h]
\centering
\includegraphics[width=7.5cm]{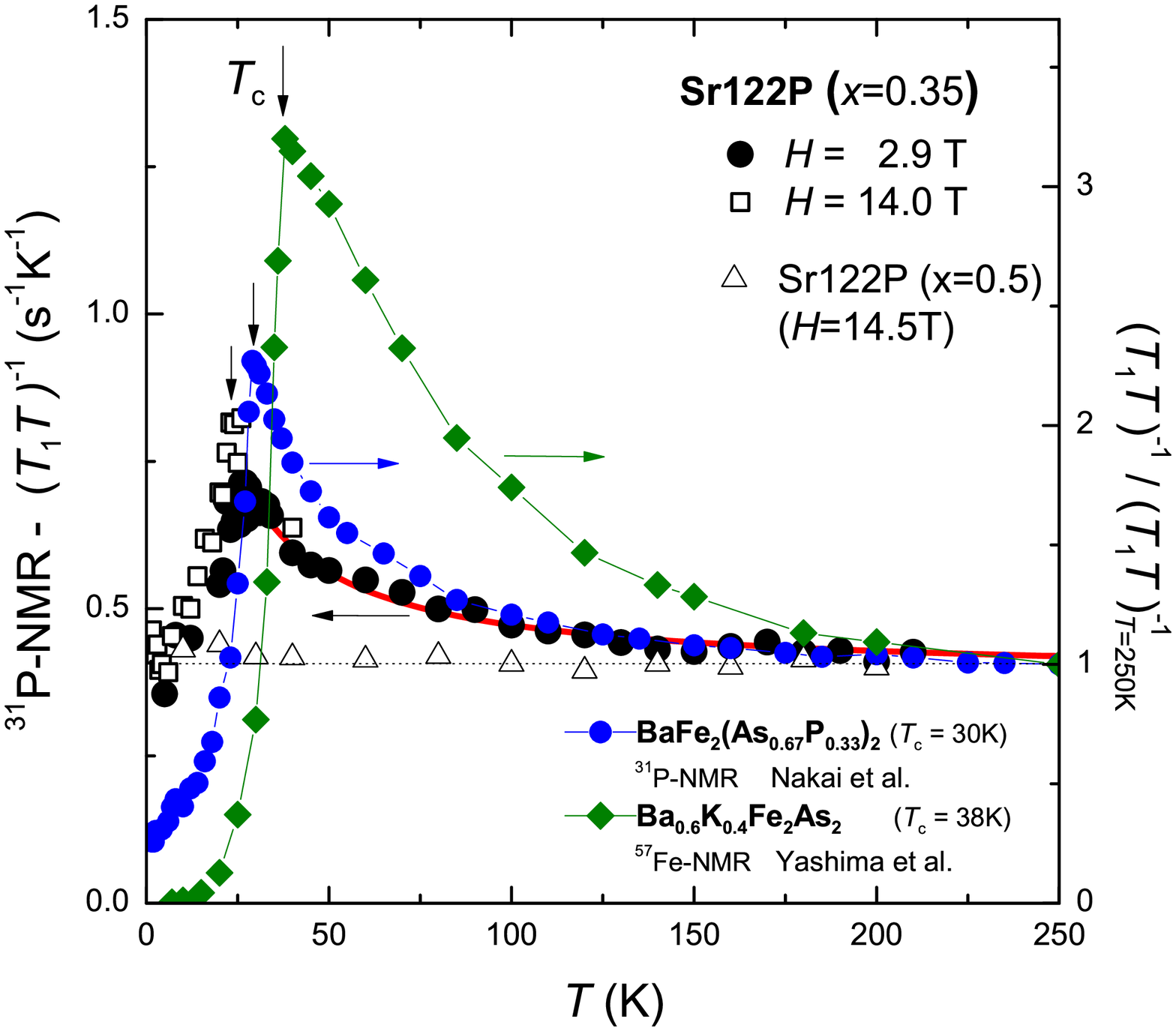}
\caption[]{(Color online)  $T$ dependence of $^{31}$P-NMR $(T_1T)^{-1}$ in the normal state for Sr122P along with the results for Ba122P with $T_c$=30 K\cite{NakaiPRL} and for BaK122 with $T_c$=38 K \cite{Yashima}. Here, the data for $^{31}$P($^{57}$Fe)-NMR $(T_1T)^{-1}$ in Ba122P(BaK122) are normalized by the value at 250 K. 
Solid curve is a fit for Sr122P($x$=0.35) on the assumption of $(T_{1}T)^{-1} = a/(T+\theta)+b$ with parameters $a$=8.7, $b$=0.38, and $\theta $=$-0.5$, indicating the closeness to the AFM QCP.
}
\label{1/T1TvsT}
\end{figure}

Nuclear spin-lattice relaxation rate $1/T_1$ for both samples was obtained from the recovery of nuclear magnetization by fitting with a simple exponential recovery curve of $m(t) = [M_0-M(t)]/M_0 = \exp(-t/T_1)$ for $^{31}P$($I=1/2$). 
Here $M_0$ and $M(t)$ are the respective nuclear magnetizations of $^{31}$P for the thermal equilibrium condition and at a time $t$ after the saturation pulse.  
Typical recovery curves of $x$=0.35 are presented for normal- and SC state in Figs. \ref{recovery}(a) and  \ref{recovery}(b), respectively. 
The $m(t)$ can be determined by a single component of $T_1$ for both $x$=0.35 and 0.5.  
Figure~\ref{1/T1TvsT} shows the  $T$ dependence of $^{31}$P-NMR $1/T_1T$ in the normal state, which increases significantly upon cooling down to $T_{\rm c}$. 
In general, $1/T_1T$ is described as, 
\[
\frac{1}{T_1T}\propto \sum_{\bm q} |A_{\bm q}|^2 \frac{\chi''({\bm q},\omega_0)}{\omega_0},
\]
where $A_{\bm q}$ is a wave-vector (${\bm q}$)-dependent hyperfine-coupling constant, $\chi({\bm q},\omega)$ a dynamical spin susceptibility, and $\omega_0$ an NMR frequency.  
Since the Knight shift does not change  in the normal state, the increase of $1/T_1T$ upon cooling is due to the development of AFM correlations as well as the case in  Ba122P\cite{NakaiPRL} and BaK122\cite{Yashima}. 
We assume two-dimensional (2D) AFM correlations model that predicts a relation of $1/T_1T\propto \chi_{\rm Q}(T) \propto 1/(T+\theta)$  when a system is close to an AFM quantum critical point (QCP)\cite{Moriya}. 
Here, the staggered susceptibility $\chi_{\rm Q}(T)$ with an AFM propagation vector ${\bm q}$=${\bm Q}$ follows a Curie-Weiss law. Since $1/T_1T$ diverges towards $T \rightarrow 0$ when $\theta=0$, $\theta$ is a measure of how close a system is to an AFM QCP.  Actually, as shown by the solid curve in Fig.~\ref{1/T1TvsT}, the $T$ dependence of $1/T_{1}T$ for $x$=0.35 can be fitted by assuming $1/T_{1}T= a/(T+\theta)+b$ with parameters $a$=8.7, $b$=0.38, and $\theta $=$-0.5$\,K. In particular, it is notable that $\theta$ is nearly zero for $x$=0.35, indicating the closeness to the AFM QCP as well as in Ba122P with $a\sim$7, $\theta\sim 0$~K, and $b\sim$0.17 \cite{NakaiPRL}.
Such a critical enhancement of $1/T_1T$ disappears in $x$=0.5, as shown in Fig.~\ref{1/T1TvsT}.

We address why the enhancement of $1/T_1T$ at low $T$ becomes distinct in going from Sr122P, Ba122P, to BaK122 as seen in Fig.~\ref{1/T1TvsT}. As listed in table \ref{table1}, the isovalent substitution of P for As largely reduces the $c$-axis length and a pnictgen height ($h_{Pn}$), whereas the $a$-axis length remains constant, which  may be equivalent to applying an uniaxial stress along the $c$-axis for (Fe$Pn$)$^{-}$ layers. 
The band-structure calculation for Ba122P has predicted that one of the hole FSs mainly derived from the $XZ/YZ/Z^2$ orbitals exhibits three dimensional (3D) features when $h_{Pn}$ becomes lower in the 122 compounds, whereas the hole FS from $X^2-Y^2$ orbital keeps a quasi-2D character\cite{Suzuki_Kuroki}. 
Although the local lattice parameters of (Fe$Pn$)$^{-}$ resemble for both Ba122P and Sr122P, the reduction of $c$-axis length is more significant for Sr122P. 
In this context, appearance of three-dimensionality in the electronic structure may cause the nesting of FSs to weaken in some of the multiple bands.

\begin{table}
\centering
\caption[]{\footnotesize  Superconducting characteristics and lattice parameters for SrFe$_2$(As$_{0.65}$P$_{0.35}$)$_2$ (Sr122P)\cite{Shi} along with those of Ba$_{0.6}$K$_{0.4}$Fe$_2$As$_2$ (BaK122) \cite{Rotter} and BaFe$_2$(As$_{0.67}$P$_{0.33}$)$_2$ (Ba122P)\cite{Kasahara,NakaiPRB}. 
}
\begin{tabular}{lccc}
\hline
           & BaK122\cite{Rotter} & Ba122P\cite{Kasahara,NakaiPRB} & Sr122P($x$=0.35)\cite{Shi} \\
\hline
$T_c$[K]           &  38     &  30      &   26   \\
Type of SC Gap     &  Full   & gapless  &  gapless  \\
\hline
$a$[\AA]           &  3.917  &  3.92    &  3.92$\dagger$ \\
$c$[\AA]           &  13.3   &  12.8    &  12.23$\dagger$ \\
$h_{Pn}$[\AA]      &  1.38   &  1.32    &  1.32$\dagger$ \\
\hline
\end{tabular}
\label{table1}
\begin{flushleft}
\footnotesize{$\dagger$) The values for $x$=0.35 are interpolated from the data in ref\cite{Shi}.}
\end{flushleft}
\end{table}

\begin{figure}[h]
\centering
\includegraphics[width=7.5cm]{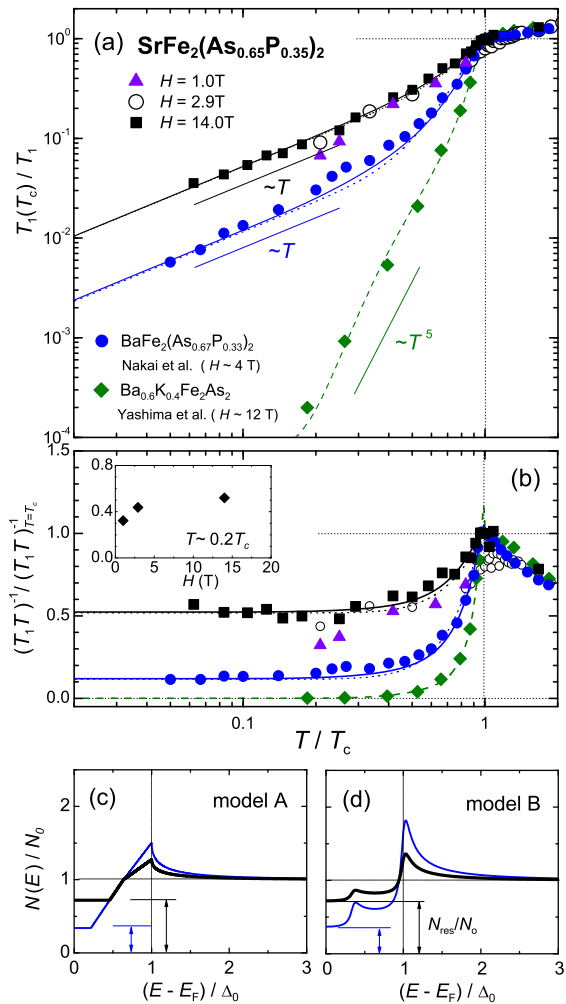}
\caption[]{(Color online) (a) Plots of $^{31}$P-NMR $T_1(T_c)/T_1$  and (b) $(T_1T)^{-1}/(T_1T)^{-1}_{T_c}$ versus $T/T_c$ for Sr122P along with those for Ba122P with $T_c$=30 K \cite{NakaiPRB} and  BaK122 with $T_c$=38 K \cite{Yashima}. Inset shows the $H$ dependence of $(T_1T)^{-1}/(T_1T)^{-1}_{T_c}$ at $T\sim0.2T_c$.
Dotted and Solid curves are simulations on the assumption of (c) DOS for simple nodal gap with RDOS (model A)  and (d) DOS for multigap composed of a small nodal gap with RDOS and full gaps (model B). In Sr122P, the model B is appropriate (see the text).}
\label{T1_SC}
\end{figure}

Next, we present the SC characteristics in Sr122P($x$=0.35) with $T_c$=26 K. 
Figure~\ref{T1_SC}(a) shows a plot of $^{31}$P-NMR $T_1(T_c)/T_1$ versus $T/T_c$ for Sr122P along with the results for Ba122P\cite{NakaiPRB} with $T_c$=30 K and BaK122\cite{Yashima} with $T_c$=38 K. 
The $1/T_1$s for Sr122P ($x$=0.35) decrease steeply below $T_c$ without a coherence peak and follow a $T$-linear dependence below $T/T_c<$0.5 when $H=14$ T. 
The $1/T_1T$ normalized by the value at $T_c$ remains a finite, as shown in Fig.~\ref{T1_SC}(b). 
Although it shows a weak $H$ dependence, it still remains a finite even in low $H$ limit at $T\sim0.2 T_c$, as seen in the inset of Fig.~\ref{T1_SC}(b), pointing to a gapless SC dominated by a large contribution of low-lying quasiparticle excitations at $E_F$. 
It is in contrast with the power-law behavior without $T$-linear dependence in $1/T_1$ in BaK122 despite of the application of high external field (12 T), which was consistently accounted for by the fully-gapped $s_{\pm}$-wave state\cite{Yashima}.
Generally, the $1/T_1T$ in the gapless SC state is related to the square of RDOS at $E_F$ ($N_{\rm res}^2$), and hence the fraction of RDOS ($N_{\rm res}/N_0$) to a normal-state DOS $N_0$ is given by 
\[
\frac{N_{\rm res}}{N_0}=\sqrt{\frac{(T_1T)^{-1}_{T\rightarrow 0}}{(T_1T)^{-1}_{T=T_c}}}. 
\]
Using this relation, it is deduced that $N_{\rm res}/N_0$ is approximately  0.72 when $H=14$ T. 
Although the $1/T_1T$ does not exhibit a constant at low temperatures when $H<$ 2.9 T, if we assume it by a value at $T\sim0.2T_c$, the $N_{\rm res}/N_0$s are as large as 0.56 and 0.66 in $H$=1 T and 2.9 T, respectively, being much larger than $N_{\rm res}/N_0\sim$0.34($H\sim$4 T) for the nodal SC in Ba122P \cite{NakaiPRB}, and $N_{\rm res}/N_0$=0 for fully gapped SC in BaK122\cite{Yashima}. 
The $H$ dependence of RDOS in this compound is discussed by specific heat measurements later. 
The solid curves in Figs. \ref{T1_SC}(a) and \ref{T1_SC}(b) are simulations when we assume two possible SC-gap models: (A) single nodal gap of $\Delta(\phi)=\Delta_0\sin(2\phi)$ with $N_{\rm res}/N_0$ (denoted as model A), which was tentatively assumed for Ba122P\cite{NakaiPRB}, and (B) multigap with the fully gapped $s_\pm$-wave state in some bands\cite{Yashima} and gapless in other bands under $H$ (model B), which reproduce the $T$ dependence of $1/T_1$ in the gapless SC state. 
The simulations based on the respective models A and B for Sr122P(Ba122P\cite{NakaiPRB}) are shown by the dotted curves with parameters of $2\Delta_0\sim 5k_BT_c$ ($6k_BT_c$)  and $N_{\rm res}/N_0$=0.72(0.34) and by the solid curves with parameters of $2\Delta_0^L\sim 5k_BT_c$ ($6k_BT_c$) for bands with full gaps and a smearing factor of DOS $\eta=0.05\Delta_0^L$ assumed in BaK122\cite{Yashima} besides the gapless band with $N_{\rm res}/N_0$=0.72(0.34). 
The assumed DOS  for the model A and B are indicated in Figs.~\ref{T1_SC}(c) and \ref{T1_SC}(d), respectively. 

\begin{figure}[h]
\centering
\includegraphics[width=7cm]{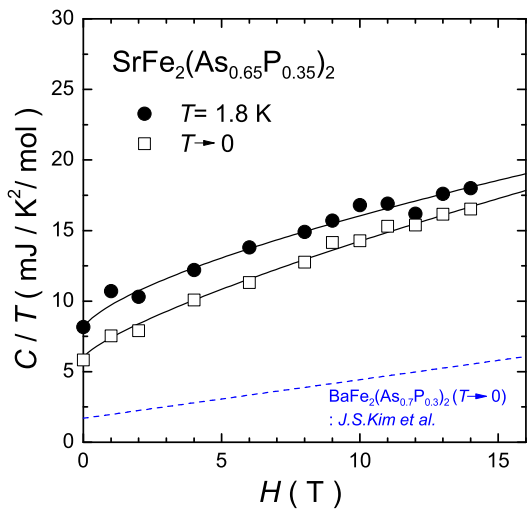}
\caption[]{Field dependence of specific heat $C(T)/T$ at 1.8 K and  the $\gamma(\equiv \lim_{T\rightarrow 0} C(T)/T)$ estimated from the $T$ dependence of $C(T)/T$ above 1.8 K\cite{Kobayashi}. 
The solid curves are a fitting with the function of $H^\alpha$($\alpha \sim 0.7$). The broken line is the $C(T)/T$  at $T\rightarrow 0$ for BaFe$_2$(As$_{0.7}$P$_{0.3}$)$_2$\cite{JSKim}. 
}
\label{C}
\end{figure}

To reveal the origin of the RDOS, we present the field dependence of the specific heat $C(T)/T$ at 1.8 K and the residual $\gamma(\equiv \lim_{T\rightarrow 0} C(T)/T)$  extrapolated to $T=0$ K by fitting with $C(T)/T=\gamma + \beta T^2 + \delta T^4$ in Fig. \ref{C}\cite{Kobayashi}. 
The result reveals that $\gamma$ is proportional to $\sim H^{0.7}$ for $0<H<$ 14 T, indicating that the low-energy quasiparticle excitation is dominated by the nodal gap. 
The significant difference from the $H^{0.5}$ dependence expected in the $d$-wave line node gap state suggests the multiband effect with the unequal sizes of gaps\cite{Bang}, in addition to the Doppler shift of the quasiparticle energy in the vortex state, as seen in Ba122P \cite{Hashimoto2,NakaiPRB,Volovik,Wang}.
Using $\gamma_n \equiv C(T_c)/T_c \sim$ 28 mJ/ (mol K$^2$) at $T_c$, the fraction of RDOS  was estimated to be $\sim60$\% of $N_0$ in the application of $H$=14 T, which is roughly consistent with the result obtained by the $^{31}$P-NMR experiment. 
Even when $H=0$ without vortices, the residual $\gamma$ that derives from the impurity scattering remains $\sim$6 mJ/(mol K$^2$) at $T$=0 K \cite{Kobayashi}, corresponding to $\gamma/\gamma_n \sim$ 0.21, which was larger than 0.11 in Ba122P\cite{JSKim,Wang}. 
Taking into account that the highly homogeneous electronic state is guaranteed by the narrow $^{31}$P-NMR linewidths, the low-energy quasiparticle DOS in the nodal gap is sensitively induced even by very small amount of impurity scattering. 
This behavior is in contrast with the case of BaK122 ($T_c=38$ K) characterized by a fully gaped SC state, where the application of field of $H= 6 \sim 12$ T does not induce such low-energy quasiparticle excitation \cite{Yashima,Fukazawa}. 
It is therefore likely that the presence of RDOS is inherent in the 122 iron-pnictides in which a lattice contraction takes place by either the chemical substitution of P for As or the application of pressure.
The results are consistently argued with the multigap feature that possess a large gap-size ratio $\Delta_L/\Delta_S$ of a large gap ($\Delta_L$) to a small gap ($\Delta_S$). 
Here, we remind the well-established cases of the multiband SC with a large gap-size ratio $\Delta_L/\Delta_S\sim$ 3, which was demonstrated in MgB$_2$\cite{MgB2} and NbSe$_2$\cite{NbSe2}. 
Even in these conventional s-wave superconductors, low-lying excitations are induced due to the delocalization of quasiparticles within the vortices at significantly low  fields at $\sim H_{c2}/10$. The rapid increase in $N_{\rm res}/N_0$ at low $H$ in these compounds is a consequence of the small magnitude of the second gap and the associated small condensation energy. 
Likewise, in Sr122P, it might be expected that the large fractional RDOS is promptly induced by fields sufficiently smaller than $H_{c2}\sim$ 70 T\cite{Kida}. 
It can be attributed to that the presence of a small gap with nodes causes the quasiparticles within the vortices to be more significantly delocalized than in conventional fully-gapped  superconductors. 
In this context, it is a multigap effect with a gap-size ratio of different bands being significantly large to induce the large contribution for the gapless low-lying quasiparticle excitations in the SC state for Sr122P.
Under these situations, we claim that the model B (Fig.~\ref{T1_SC}(d)), which deals with the multigap system, is appropriate for Sr122P, although both models seem to be able to reproduce the experimental results. 
Such multiband scenario gives a reasonable explanation for the fact that $T_c=26$ K of Sr122P is still relatively high irrespective of the gapless SC state in some of the multiple bands, which may be characteristic for P-doped 122 compounds. 



Finally, in order to shed light on an origin of the small gap with nodes, we compare the SC characteristics and the lattice parameters for Sr122P, Ba122P and BaK122 in Table \ref{table1}.
The fully-gapped $s_{\pm}$-wave state in BaK122 is characterized by the largest $c$-axis length and the highest $h_{Pn}$, whereas the gapless SC state in Sr122P is by the lowest $h_{Pn}$ and the reduction of $c$-axis length, which is more significant than in others. 
Actually, the residual DOS was more prominent in Sr122P than in Ba122P\cite{JSKim,Wang,Kobayashi,NakaiPRB}, as seen via both  the NMR and the specific heat measurements, implying that the portion of the Fermi surfaces with RDOS becomes large in going from Ba122P to Sr122P. 
According to the band-structure calculation for Ba122P,  the orbital character of one of the hole sheets is very sensitive to $h_{Pn}$ when As is substituted by P in 122 compounds, and pointed out that 3D nodal structures appear in a largely-warped hole FS having a strong $Z^2/XZ/YZ$ orbital character in Ba122P\cite{Suzuki_Kuroki}, which was supported experimentally by ARPES\cite{Yoshida}. 
Since more prominent 3D feature in electronic structure may be anticipated in Sr122P than in Ba122P, as corroborated by the normal-state properties discussed above, it is possible that the origin of a nodal-structure in some of the multiple bands is ascribed to the largely-warped Fermi surface. 
In particular, the important point on the scenario is that the fully-gapped $s_{\pm}$-wave state in the other quasi-2D bands keeps $T_c$ still high through the interband scattering between quasi-2D electron and hole FSs within the $k_x$-$k_y$ plane in association with the multiband SC nature\cite{Suzuki_Kuroki}.
In other context, the monotonous decrease of the maximum $T_c$ from BaK122, Ba122P, Sr122P, to Ca122P is probably due to the decrease of the total SC condensation energy for the fully-gapped $s_{\pm}$-wave  state due to the worse nesting properties in some of the multiple bands.
Another possible scenario for  a nodal gap structure is the closed nodal loops located at the flat parts of the electron Fermi surface, which was proposed in Ba122P by Yamashita {\it et al.}\cite{Yamashita}. 
To reveal the strong material dependence of SC-gap structure in Fe-based superconductors, the further systematic studies on 122 compounds are desired to focus on the relation between the local structure of Fe$P_n$ layer and the gap structure.


In  summary, the $^{31}$P-NMR and specific heat measurements on single crystallines SrFe$_2$(As$_{0.65}$P$_{0.35}$)$_2$ (Sr122P) have revealed an unconventional gapless SC with nodes, which is accounted for by the multiband feature with the gap-size ratio of different bands being significantly large. 
The normal-state property is dominated by the development of antiferromagnetic correlations  pointing to the closeness to an antiferromagnetic quantum critical point for $x$=0.35. 
However, the comparison between the other 122 compounds reveals the reduction of AFM correlations in Sr122P, implying the worse nesting property in some of multiple bands due to three-dimensional feature, which is suggested in Ba122P by the band-calculation and ARPES results\cite{Suzuki_Kuroki,Yoshida}: The three-dimensionality appears in the largely warped hole Fermi surface, whereas the other quasi-two dimensional electron and hole Fermi surfaces within the $k_x$-$k_y$ plane is not significantly modified. 
If we assume the nodal gap on the largely warped hole Fermi surface, this scenario gives a reasonable explanation for relatively high $T_c$ regardless of the gapless feature in this compound, where the interband scattering within the quasi 2D bands stabilizes the fully-gapped $s_{\pm}$-wave state. 
The present results in association with the multiband nature will develop an insight into the strong material dependence of SC-gap structure in Fe-based superconductors.


{\footnotesize 
We thank K. Kuroki, S. Kawasaki, T. Yoshida, and K. Ishida for valuable discussion. This work was supported by a Grant-in-Aid for Specially Promoted Research (20001004) and by the Global COE Program (Core Research and Engineering of Advanced Materials-Interdisciplinary Education Center for Materials Science) from the Ministry of Education, Culture, Sports, Science and Technology (MEXT), Japan.
}


\begin{thebibliography}{99} 

\bibitem{Rotter} M. Rotter, M. Tegel, and D. Johrendt, Phys. Rev. Lett. {\bf 101},  107006 (2008).
\bibitem{Hashimoto1} K. Hashimoto, T. Shibauchi, T.Kato, K. Ikada, R. Okazaki, H. Shishido, M. Ishikado, H. Kito, A. Iyo, H. Eisaki, S. Shamoto, and Y. Matsuda, Phys. Rev. Lett. {\bf 102}, 017002 (2009).
\bibitem{Ding} H. Ding, P. Richard, K. Nakayama, K. Sugawara, T. Arakane, Y. Sekiba, A. Takayama, S. Souma, T. Sato, T. Takahashi, Z. Wang, X. Dai, Z. Fang, G. F. Chen, J. L. Luo, and N. L. Wang, Europhys. Lett. {\bf 83}, 47001 (2008).
\bibitem{Yashima} M.~Yashima, H.~Nishimura, H.~Mukuda, Y.~Kitaoka, K.~Miyazawa, P. M.~Shirage, K.~Kiho, H.~Kito, H.~Eisaki, and A.~Iyo, J. Phys. Soc. Jpn. {\bf 78}, 103702 (2009).
\bibitem{Xu} Y. M. Xu, Y. B. Huang, X. Y. Cui, E. Razzoli, M. Radovic, M. Shi, G. F. Chen, P. Zheng, N. L. Wang, C. L. Zhang, P. C. Dai, J. P. Hu, Z. Wang, and H. Ding, Nature Phys. {\bf 7}, 198 (2011). 
\bibitem{Zheng} Z. Li, D. L. Sun, C. T. Lin, Y. H. Su, J. P. Hu, and G. Q. Zheng, Phys. Rev. B {\bf 83}, 140506(R) (2011). 

\bibitem{Kasahara} S. Kasahara, T. Shibauchi, K. Hashimoto, K. Ikada, S. Tonegawa,  R. Okazaki, H. Shishido, H. Ikeda, H. Takeya, K. Hirata, T. Terashima, and Y. Matsuda, Phys. Rev. B {\bf 81}, 184519 (2010).
\bibitem{NakaiPRB} Y.~Nakai, T.~Iye, S.~Kitagawa, K.~Ishida, S.~Kasahara,  T.~Shibauchi, Y.~Matsuda, and T.~Terashima, Phys. Rev. B {\bf 81}, 020503(R) (2010).
\bibitem{Hashimoto2} K. Hashimoto, M. Yamashita, S. Kasahara, Y. Senshu, N. Nakata, S. Tonegawa, K. Ikada, A. Serafin, A. Carrington, T. Terashima, H. Ikeda, T. Shibauchi, and Y. Matsuda, Phys. Rev. B  {\bf 81}, 220501 (2010).
\bibitem{JSKim} J. S. Kim, P. J. Hirschfeld, G. R. Stewart, S. Kasahara, T. Shibauchi, T. Terashima, and Y. Matsuda, Phys. Rev. B  {\bf 81}, 214507 (2010).
\bibitem{Wang} T. Y. Wang, J. S. Kim, G. R. Stewart, P. J. Hirschfeld, S. Graser, S. Kasahara, T. Terashima, Y. Matsuda, T. Shibauchi, I. Vekhter, Phys. Rev. B {\bf 84}, 184524 (2011).
\bibitem{Yamashita} M. Yamashita, Y. Senshu, T. Shibauchi, S. Kasahara, K. Hashimoto, D. Watanabe, H. Ikeda, T. Terashima, I. Vekhter, A.B. Vorontsov, and Y. Matsuda, Phys. Rev. B {\bf 84}, 060507(R) (2011).
\bibitem{Kitagawa} K. Kitagawa, N. Katayama, H. Gotou, T. Yagi, K. Ohgushi, T. Matsumoto, Y. Uwatoko, and M. Takigawa, Phys. Rev. Lett. {\bf 103}, 257002 (2009). 
\bibitem{Shimojima} T. Shimojima, F. Sakaguchi, K. Ishizaka, Y. Ishida, T. Kiss, M. Okawa, T. Togashi, C.-T. Chen, S. Watanabe, M. Arita, K. Shimada, H. Namatame, M. Taniguchi, K. Ohgushi, S. Kasahara, T. Terashima, T. Shibauchi, Y. Matsuda, A. Chainani, and S. Shin, Science {\bf 332}, 564 (2011). 
\bibitem{Feng} Y. Zhang, Z. R. Ye, Q. Q. Ge, F. Chen, Juan Jiang, M. Xu, B. P. Xie, and D. L. Feng, (arXiv.1109.0229).
\bibitem{Suzuki_Kuroki} K. Suzuki, H. Usui, and K. Kuroki, J. Phys. Soc. Jpn. {\bf 80}, 013710 (2011).
\bibitem{Yoshida} T. Yoshida, I. Nishi, S. Ideta, A. Fujimori, M. Kubota, K. Ono, S. Kasahara, T. Shibauchi, T. Terashima, Y. Matsuda, H. Ikeda, and R. Arita, Phys. Rev. Lett. {\bf 106}, 117001 (2011). 
\bibitem{Shi} H. L. Shi, H. X. Yang, H. F. Tian, J. B. Lu, Z. W. Wang, Y. B. Qin, Y. J. Song, and J. Q. Li, J. Phys:Condens. Matter {\bf 22}, 125702 (2010).
\bibitem{Kasahara2} S. Kasahara, T. Shibauchi, K. Hashimoto, Y. Nakai, H. Ikeda, T. Terashima, and Y. Matsuda, Phys. Rev. B {\bf 83}, 060505 (2011).
\bibitem{Kobayashi} T. Kobayashi, S. Miyasaka, and S. Tajima, unpublished. 
\bibitem{NakaiPRL} Y.~Nakai, T.~Iye, S.~Kitagawa, K.~Ishida, H.~Ikeda, S.~Kasahara, H.~Shishido, T.~Shibauchi, Y.~Matsuda, and T.~Terashima, Phys. Rev. Lett. {\bf 105}, 107003 (2010). 
\bibitem{Moriya} T. Moriya and K. Ueda, Adv. Phys. {\bf 49}, 555 (2000).
\bibitem{Volovik} G. E. Volovik, JETP Lett. {\bf 58}, 469 (1993).
\bibitem{Bang} Y. Bang, Phys. Rev. Lett. {\bf 104}, 217001 (2010).
\bibitem{Fukazawa} H. Fukazawa, T. Yamazaki, K. Kondo, Y. Kohori, N. Takeshita, P. M. Shirage, K. Kihou, K. Miyazawa, H. Kito, H. Eisaki, and A. Iyo, J. Phys. Soc. Jpn. {\bf 78}, 033704 (2009).
\bibitem{MgB2} F. Bouquet, R. A. Fisher, N. E. Phillips, D. G. Hinks, and J. D. Jorgensen, Phys. Rev. Lett. {\bf 87}, 047001 (2001).
\bibitem{NbSe2} E. Boaknin, M. A. Tanatar, J. Paglione, D. Hawthorn, F. Ronning, R.W. Hill, M. Sutherland, L. Taillefer, J. Sonier, S.M. Hayden, and J.W. Brill, Phys. Rev. Lett. {\bf 90}, 117003 (2003).
\bibitem{Kida} T. Kida {\it et al.}, in preparation. 


\end{thebibliography}
\end{document}